\documentclass[fleqn,12pt,twoside]{article}
\usepackage{espcrc1}

\usepackage{graphicx}
\usepackage[figuresright]{rotating}

\newcommand{\AmS}{{\protect\the\textfont2
  A\kern-.1667em\lower.5ex\hbox{M}\kern-.125emS}}

\def\be{\begin{equation}}
\def\ee{\end{equation}}
\def\bea{\begin{eqnarray}}
\def\eea{\end{eqnarray}}
\def\oneh{{\textstyle {1\over 2}}}

\def          
\simeq{{\ \lower2pt\hbox{$-$}\mkern-13mu \raise2pt \hbox{$\sim$}\ }}

\title{Generalized parton distributions of the nucleon in constituent quark
       models}

\author{S. Boffi\address[DIP]{Dipartimento di Fisica Nucleare e Teorica,
Universit\`a degli Studi di Pavia \\ 
        and INFN, Sezione di Pavia, Pavia, Italy},%
        B. Pasquini\addressmark[DIP]$^,$
	\address[ECT]{ECT$^*$, Villazzano (Trento), Italy},  
        M. Traini\addressmark[ECT]$^,$
	\address{Dipartimento di Fisica, Universit\`a degli Studi di
	Trento, Povo (Trento), \\
	and INFN, Gruppo Collegato di Trento, Trento, Italy}}
       
\begin{document}

\maketitle

\begin{abstract}
Generalized parton distributions (GPDs) are studied at the hadronic
(nonperturbative) scale within  different assumptions based on a relativistic
constituent quark model. In particular, by means of a meson-cloud model we
investigate the role of nonperturbative antiquark degrees of freedom and the 
valence quark contribution. A QCD evolution of the obtained GPDs is used to 
add perturbative effects and to investigate the GPDs' sensitivity 
to the nonperturbative ingredients of the calculation at larger 
(experimental) scale.
\end{abstract}

\section{Modeling GPDs with double distributions}
We will concentrate our attention on the
chiral even (helicity conserving) distribution $H^q(\overline x,\xi,Q^2,t)$ for
partons of flavor $q$ at the hadronic scale where the models we are going to
discuss are assumed to be valid to evaluate the twist-two amplitude. 
The invariant momentum square is $t=\Delta^2=({P'}^\mu-P^\mu)^2$, $\overline x$
is the quark light-cone momentum fraction with respect to the average nucleon
momentum $\overline P^\mu = \oneh(P^\mu + {P'}^\mu)$, and the skewedness $\xi$
describes the longitudinal change of the nucleon momentum,
$2\xi=-\Delta^+/\overline P^+$. In the following the dependence on the scale
$Q^2$ is understood.

We also introduce non-singlet (valence) and
singlet quark distributions, 
\be  
\label{eq:a12}   
H^{NS} (\overline x, \xi,t)  \equiv  
\sum_q\left [H^q (\overline x, \xi,t)  +  H^q (-\overline x, \xi,t) \right ] 
=  H^{NS} (-\overline x, \xi,t), 
\ee
\be 
H^S (\overline x, \xi,t)  \equiv  
\sum_q \left [H^q (\overline x, \xi,t) -  H^q (-\overline x, \xi,t)  \right ] 
=  - H^S (-\overline x, \xi,t),  
\ee 
respectively. Besides being symmetric or antisymmetric in $\overline x$, they
are also symmetric under $\xi\to-\xi$ due to the polynomiality
property~\cite{jig}. The analogous GPD for gluons is symmetric in $\overline x$,
i.e. $H^g(\overline x,\xi,t) = H^g(-\overline x,\xi,t)$, and reduces to the
gluon density $g(x)$ in the forward limit ($\overline x\to x$, $H^g(x,0,0) =
x\,g(x)$, $x>0$). 

Following ref.~\cite{Rad99} we will assume a factorized $t$ dependence
determined by some form factor, and parametrize the $t$-dependent part in terms
of double distributions (DDs) involving a given profile function and the forward
parton distribution $q(x)$ derived in some model. 
At the hadronic scale $Q_0^2$, where
the short range (perturbative) part of the interaction is negligible and,
therefore, the glue and sea are suppressed, the long range (confining) part of
the interaction produces a proton composed by (three) valence quarks,
mainly~\cite{PaPe}. Therefore quark models are suitable
to construct the parton distribution  at the scale $Q_0^2$~\cite{JaRo}.

According to the method discussed in two recent papers~\cite{BPT03,BPT04} along
the lines of ref.~\cite{diehl2}, we evaluate the valence contribution
$q^{bare}(x)$ to the parton distribution
 within relativistic light-front constituent quark models (CQMs) at the scale
$Q_0^2$. This distribution automatically fulfills the support condition and
satisfies the (particle) baryon number and momentum sum rules. Results presented
here are obtained for the hypercentral CQM~\cite{FTV99}.

\section{Parton distributions and the meson-cloud model}
The meson-cloud model introduces the sea by incorporating $q \bar q$
contributions into the valence-quark model of the parton distribution discussed
in the previous section. 
The basic hypothesis of the meson-cloud model is that the physical nucleon state
can be expanded (in the infinite momentum frame (IMF) and in the one-meson
approximation) in a series involving bare nucleons and two-particle,
meson-baryon states~\cite{m-c}. We will consider fluctuations of the proton
including both pion-nucleon and pion-Delta states. The partonic  
content of the $\Delta$ and the pion will be consistently evaluated
within the same scheme assuming light-front dynamics and valence contributions 
only. 

The quark distributions in a physical proton are then given by
\begin{equation}
q_{p} (x) = Z q_{p}^{bare}(x) + \delta q_{p}(x) , 
\label{eq:partondis}
\end{equation}
where $\delta q_{p}(x)$ is the contribution coming from the meson-baryon
fluctuations, and $Z$ is a suitable renormalization constant.

Matching the sea, valence and gluon distributions within the radiative parton
model we start evolution with continuous functions all over the range
$-1\le\overline x\le 1$ and identify the matching scale $Q_0^2=0.27$ GeV$^2$
consistent with QCD evolution equations~\cite{GRV9295,TVMZ97}. 

\section{Results and discussion}
Results are presented in this section according to the model based on DDs. The
$t$ dependence is dropped from the very beginning and could be reintroduced in
the final results by an appropriate $t$-dependent factor. The D-term is omitted
as well. 
The QCD evolution was numerically performed to NLO accuracy according to a
modified version of the code of ref.~\cite{FmcD02b} (see~\cite{PTB} for further
details and results). 

The singlet quark, non-singlet quark and gluon GPDs obtained in the model have
been studied as a function of $\overline x$, $\xi$ and $Q^2$. With no initial
gluons and an input parton distribution given by $q^{bare}(x)$ the model already
gives a nonvanishing contribution to quark GPDs in the ERBL region $\vert
\overline x\vert<\xi$ at the
hadronic scale without introducing discontinuities at $\overline x =\xi$ and
with a weak $\xi$ dependence. In particular, the absence of the sea contribution
gives $H^S= H^{NS}$ at $\overline x>\xi$. After evolution up to $Q^2=5$ GeV$^2$
GPDs are almost confined into the ERBL region with a significant $\xi$
dependence.

\begin{figure}
\centering{
\includegraphics[width=20pc,clip,trim=0 0 0 0]
{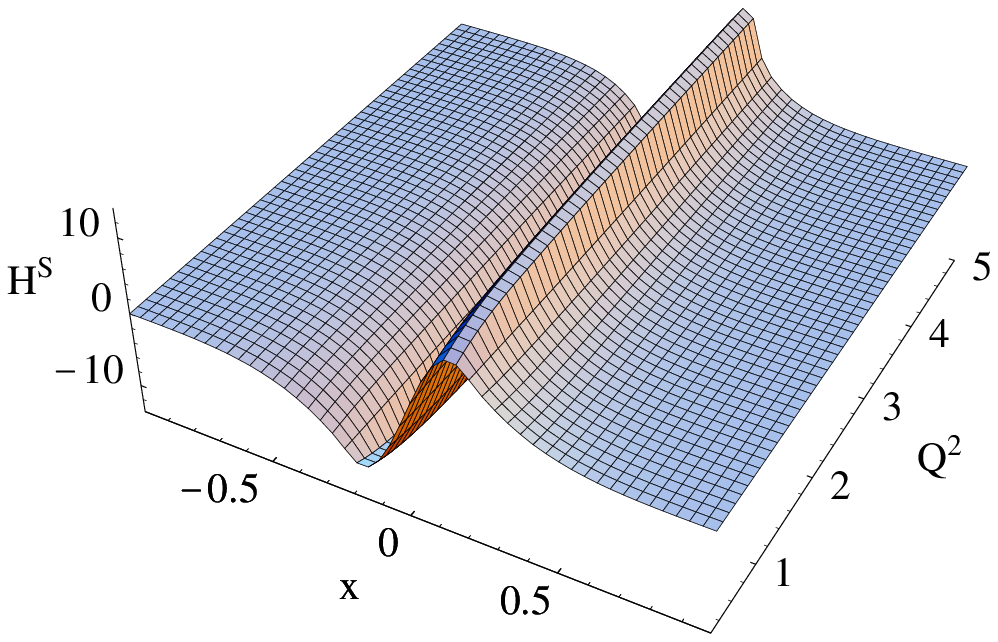}
{singlet}
\includegraphics[width=20pc,clip,trim=0 0 0 0]
{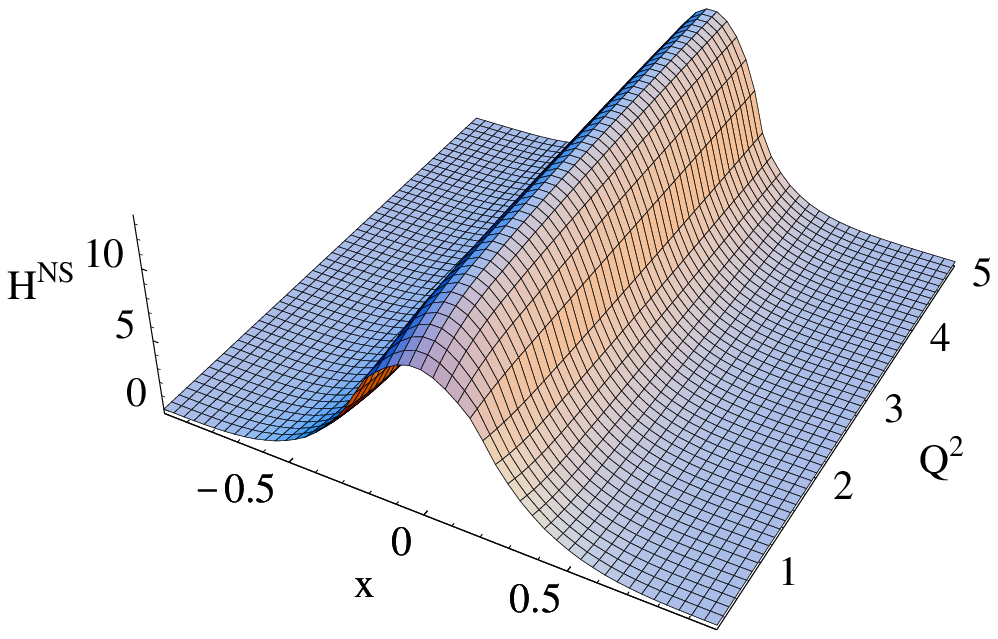}
{non-singlet}
\includegraphics[width=20pc,clip,trim=0 0 0 0]
{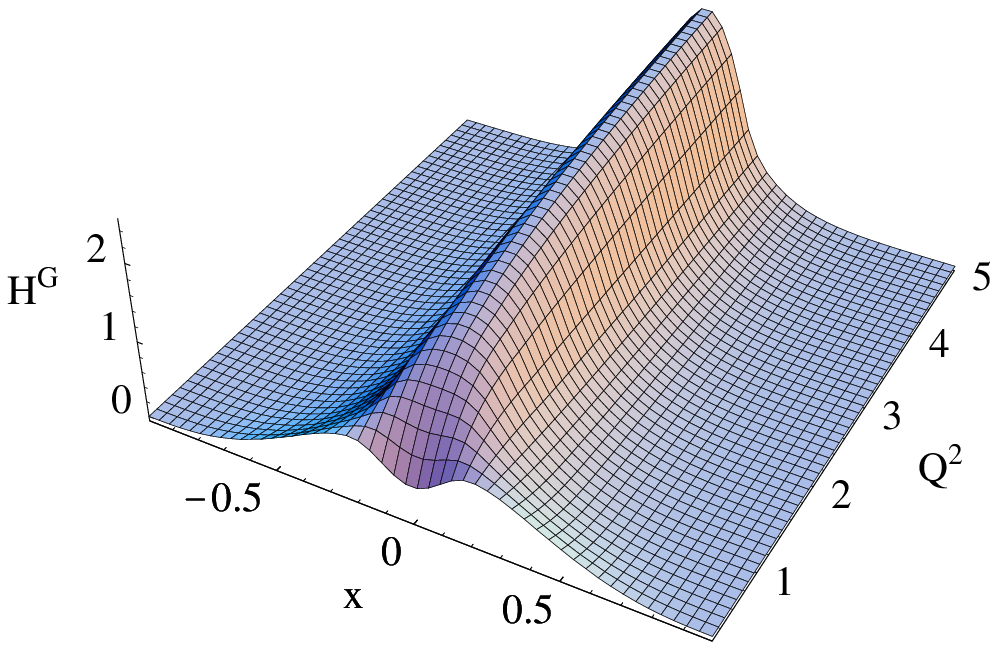}
{gluon}
\caption{Singlet and non-singlet quark and gluon GPDs at $\xi=0.2$ obtained from
NLO evolution within the double-distribution model using parton distributions of
the hypercentral CQM with the sea contribution at the initial scale $Q_0^2=0.27$
GeV$^2$.}
\label{fig:figuno}
}
\end{figure}

In fig.~\ref{fig:figuno} results are shown for QCD evolution up to $Q^2=5$
GeV$^2$ of the singlet quark, non-singlet quark and gluon GPDs obtained with the
parton distributions of the hypercentral CQM and including the sea contribution
at the initial scale $Q_0^2=0.27$ GeV$^2$. The distributions are plotted at
$\xi=0.2$ as a function of $Q^2$. In fact, the largest effects of evolution
modify the input GPDs within the first few GeV$^2$ in the $Q^2$ evolution, $Q^2=5$
GeV$^2$ being a value where GPDs have already reached a stable configuration
with respect to their asymptotic shape. 

The model already gives a nonvanishing gluon contribution at
the hadronic scale. Under evolution as the resolution scale increases the
distributions are again swept from the DGLAP domain to lie almost fully within
the ERBL region. This is a consequence of the fact that functions
with support entirely in the time-like ERBL region 
are never pushed out of the ERBL domain. In fact, the evolution in the ERBL
region depends on the DGLAP region, whereas the DGLAP evolution is independent
of the ERBL region. The qualitative result after evolution is similar to the
case without the sea~\cite{PTB} in spite of a more pronounced $\xi$ dependence of $H^S$ at
the input hadronic scale and a shape of $H^{NS}$ sensitive to the input at
$\overline x=0$.

The present results focus on the ERBL region as the most interesting one to look
at the nonperturbative effects surviving evolution. This is suggesting that one
has to investigate suitable processes under appropriate kinematic conditions to
study such effects.

This research is part of the EU Integrated Infrastructure Initiative
Hadronphysics Project under contract number RII3-CT-2004-506078 and was
partially supported by the Italian MIUR through the PRIN Theoretical Physics of
the Nucleus and the Many-Body Sytems.

\end{document}